\documentclass[a4paper,11pt]{article}
\usepackage{pos}
\usepackage{wrapfig}

\title{Reconstructing Neutrino Energy using CNNs for GeV Scale IceCube Events}
 \ShortTitle{Reconstructing Neutrino Energy using CNNs}

\author{The IceCube Collaboration \\{\normalsize \normalfont(a complete list of authors can be found at the end of the proceedings)}}

\emailAdd{micall12@msu.edu}

\abstract{Measurements of neutrinos at and below 10 GeV provide unique constraints of neutrino oscillation parameters as well as probes of potential Non-Standard Interactions (NSI). The IceCube Neutrino Observatory’s DeepCore array is designed to detect neutrinos down to GeV energies. IceCube has built the world's largest data set of neutrinos >10 GeV, making searches for NSI a computational challenge. This work describes the use of convolutional neural networks (CNNs) to improve the energy reconstruction resolution and speed of reconstructing $\mathcal{O}(10~\mathrm{GeV})$ neutrino events in IceCube. Compared to current likelihood-based methods which take seconds to minutes, the CNN is expected to provide approximately a factor of 2 improvement in energy resolution while reducing the reconstruction time per event to milliseconds, which is essential for processing large datasets.

\vspace{4mm}
{\bfseries Corresponding author:}
Jessie Micallef$^{1*}$\\
{$^{1}$ \itshape Michigan State University}\\
$^*$ Presenter

\FullConference{37$^{\rm{th}}$ International Cosmic Ray Conference (ICRC 2021)\\
		July 12th -- 23rd, 2021\\
		Online -- Berlin, Germany}
}

\begin{document}
\maketitle

\section{Introduction}

The past two decades have seen exciting advancements in both fundamental physics and astrophysics with neutrinos. GeV-scale measurements have demonstrated neutrino oscillations and nonzero neutrino masses \cite{SK-Osc,SNO-Osc}, while observations above 1 TeV indicate the existence of a population of high-energy neutrinos of astrophysical origin \cite{PhysRevLett.115.081102,Science-TXS}. However, neutrinos with energies between 10 GeV and 1 TeV remain relatively unexplored. Measurements at these energies are nontrivial, since 10 GeV is beyond the reach of accelerator and reactor neutrino experiments. To detect atmospheric neutrinos above 10 GeV, detectors must be dense enough to reconstruct the neutrino interactions while also being large enough to compensate for the steeply falling atmospheric neutrino spectrum. This is a considerable experimental challenge that IceCube is uniquely suited to handle.

\begin{wrapfigure}{r}{0.6\textwidth}
\centering
\includegraphics[scale=0.2]{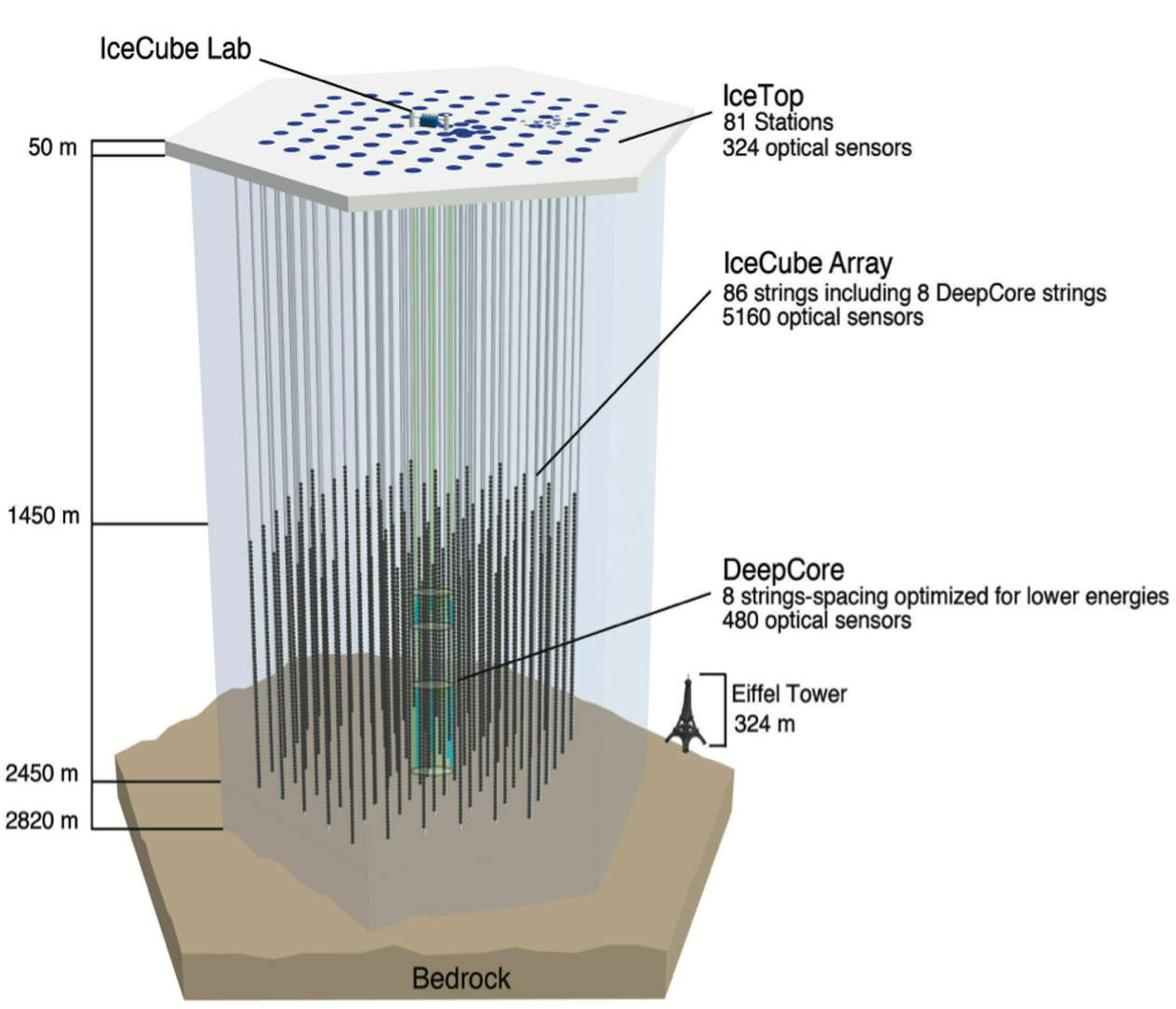}
\caption{The IceCube Detector.}
\label{fig:Detector}
\end{wrapfigure}

IceCube, located at the South Pole, instruments a cubic kilometer of ice between depths of 1450 m and 2450 m \cite{Aartsen:2016nxy} . It relies on the optical detection of Cherenkov radiation, emitted by charged particles produced from neutrino interactions in the ice or nearby bedrock, using 5160 Digital Optical Modules (DOMs). The DOMs are attached on cables called "strings", which are arranged in a hexagonal array with $\sim100$~m horizontal spacing, with 60 DOMs per string vertically separated by 17 m (Figure \ref{fig:Detector}). The DeepCore subarray includes 8 densely instrumented strings, separated horizontally by $\sim72$~m spacing and with 7 m of vertical space between the DOMs, optimized for energies as low as about 10 GeV \cite{DeepCore}. Utilizing DeepCore to detect 10 GeV neutrinos can provide us with a window into a largely unexplored regime to study the oscillation of atmospheric neutrinos \cite{IceCube-Osc}. Neutrino oscillation probabilities are a function of the ratio of the distance traveled by the neutrino to its energy, $L/E$. Therefore, improving the energy reconstruction resolution in IceCube can directly advance the ability to constrain neutrino oscillation parameters.

\section{Improving the Energy Reconstruction with a Convolutional Neural Network}

\begin{figure}[h]
\center
\begin{minipage}{.6\textwidth}
\center{\includegraphics[scale=0.45]{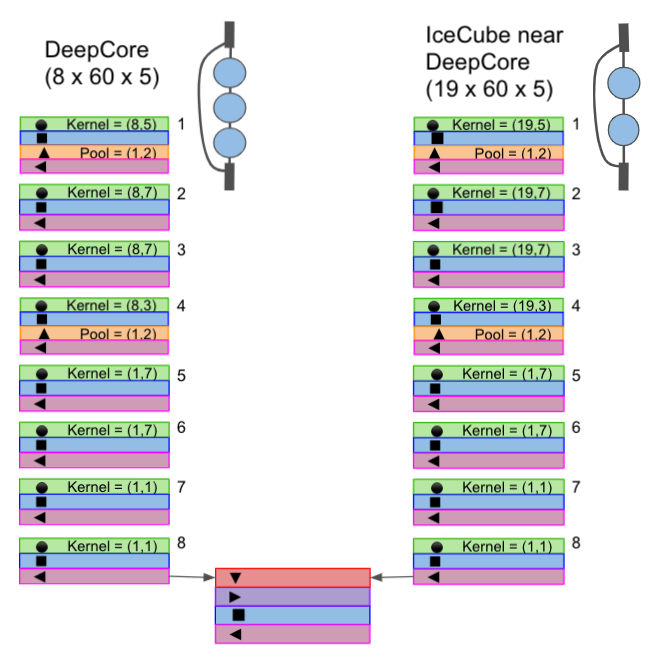}}
\end{minipage}\hspace{4em}%
\begin{minipage}{.2\textwidth}
 \center{\includegraphics[scale=0.4]{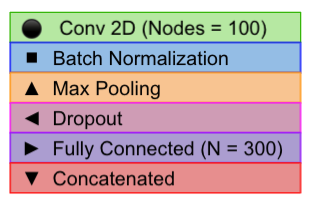}}
\end{minipage}
\caption{Detailed architecture for the Convolutional Neural Network, legend for color layers on the right.}
\label{fig:Architecture}
\end{figure}

The IceCube DOMs record the Cherenkov radiation from neutrino interactions, which can happen anywhere in the detector volume, meaning the events are characterized by translational invariance. This property allows us to apply Convolutional Neural Networks (CNNs) to the event reconstruction. CNNs are often used in image recognition as they are able to identify an object independently of where the object is positioned within an image. \cite{Krizhevsky:cnn, kaudererabrams2017quantifying}. In a CNN, a set of filters or 'kernels' is used to build maps of spatial features from the input data. By combining several convolutional layers, it is possible to reconstruct complex images. In IceCube, this architecture can be used to build a reconstruction using a network of convolutional layers (Figure \ref{fig:Architecture}). This CNN contains 8 convolutional layers which are split into two identical branches applied separately on the 8 DeepCore strings and 19 neighboring IceCube strings, since their DOMs have different vertical spacing. The two convolutional branches are then combined to train one fully connected layer on all 27 string inputs to give the final energy prediction. 

This CNN architecture is similar to the CNN applied for IceCube's high energy reconstruction \cite{Huennefeld:2017tT}. In IceCube, low energies are particularly difficult to reconstruct due to the sparseness of the detector. Low-energy events also produce less scattered light, which reduces the information available for reconstruction and makes the event topologies less distinctive. Thus, the low-energy CNN includes key optimizations from the high-energy CNN reconstruction.

\begin{figure}[h]
\center
\begin{minipage}{.4\textwidth}
\center{\includegraphics[scale=0.2]{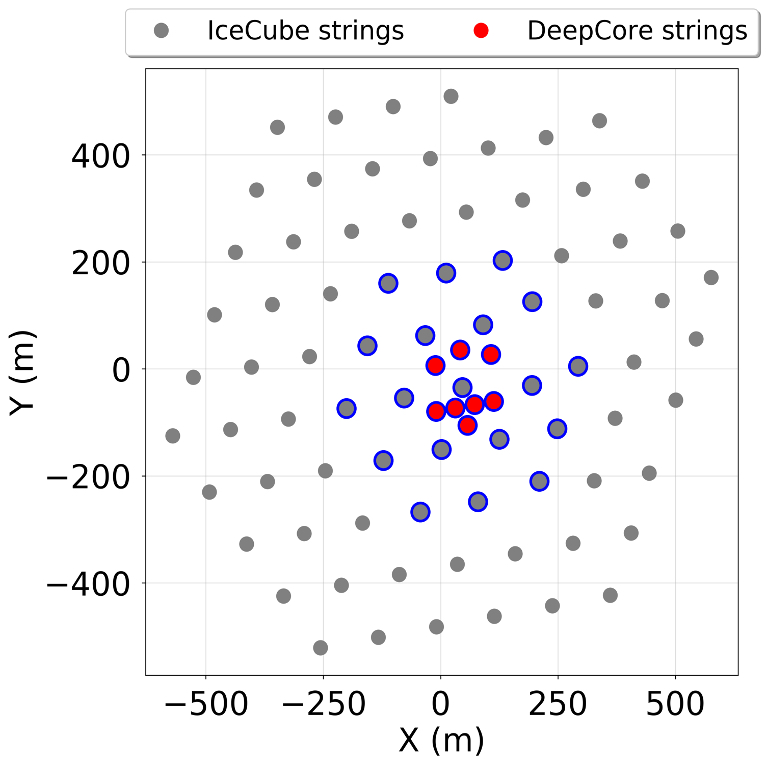}
\caption{Top view of the IceCube detector strings. DeepCore shown in red. Strings highlighted in blue are used for CNN.}
\label{fig:TopView}}\end{minipage}\hspace{4em}%
\begin{minipage}{.4\textwidth}
 \center{\includegraphics[scale=0.5]{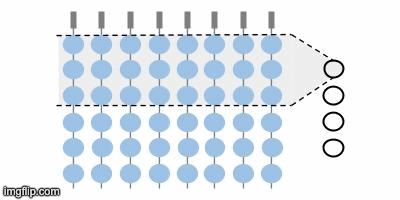}
\caption{Low energy CNN kernel spans the z-direction only, traveling down the DOMs on a string.}
\label{fig:Kernel}}
\end{minipage}
\end{figure}

The low-energy CNN uses only the DeepCore strings and the center-most IceCube strings  (Figure \ref{fig:TopView}). This accounts for the first dimension in the input array, which corresponds to the string index. All 8 DeepCore strings and the center-most 19 IceCube strings are used as input. The CNN kernel spans the vertical, or $z$-depth, of the strings only (Figure  \ref{fig:Kernel}); no convolution is applied in the $xy$-plane since the DeepCore strings are deployed in an irregular array. The second dimension in the input array uses all 60 DOMs on both the DeepCore and IceCube strings. The last dimension corresponds to the 5 input variables that summarize the (potential) multiple hits that a DOM would record in an event. These are the sum of the charge, time of the first hit, time of the last hit, charge weighted mean of the hits, and charge weighted standard deviation of the hits.

\section{Training Sample and Data Preparation}

A sample of more than 6 million unweighted, charge-current (CC) $\nu_\mu$ Monte Carlo (MC) events was generated in GENIE \cite{GENIE}, a neutrino MC generator with an emphasis on neutrino interaction physics at the few-GeV energy range, specifically for training the CNN. Of that sample, 80\% of the events ($\sim5$~million events) are used to train the CNN and 20\% ($\sim1$~million events) are used during training as a validation set to ensure the network does not overtrain, or overfit, the training sample. A potential problem when training the CNN on the atmospheric neutrino sample is that it will overtrain on low-energy showers, since the neutrino spectrum is steeply falling. To reduce such biases in the reconstruction, the MC training sample uses a flat distribution in energy between 1 and 200 GeV. The training sample is extended past the target reconstruction region up to 500 GeV so that the CNN learns to reconstruct higher energy events.

\begin{figure}[h]
\center
\begin{minipage}{.4\textwidth}
\center{\includegraphics[scale=0.28]{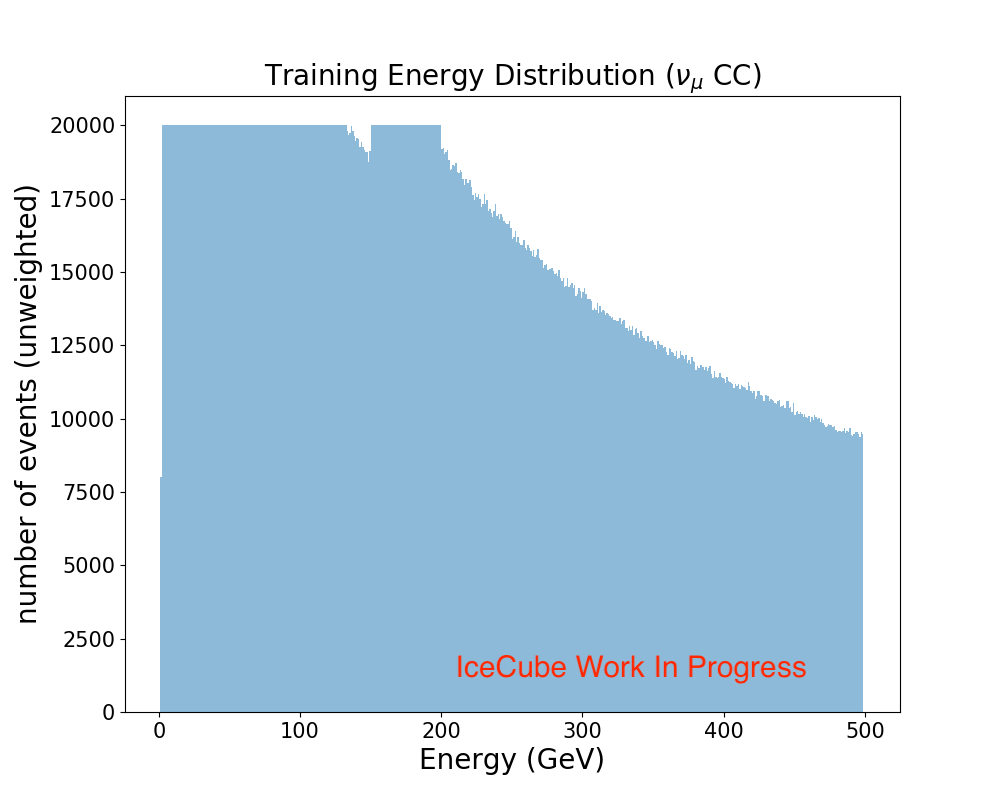}
\caption{The current training sample for the CNN, mostly uniform between the target reconstructed energies of 1-200 GeV.}
\label{fig:TrainingSample}}\end{minipage}\hspace{3em}%
\begin{minipage}{.4\textwidth}
 \center{\includegraphics[scale=0.28]{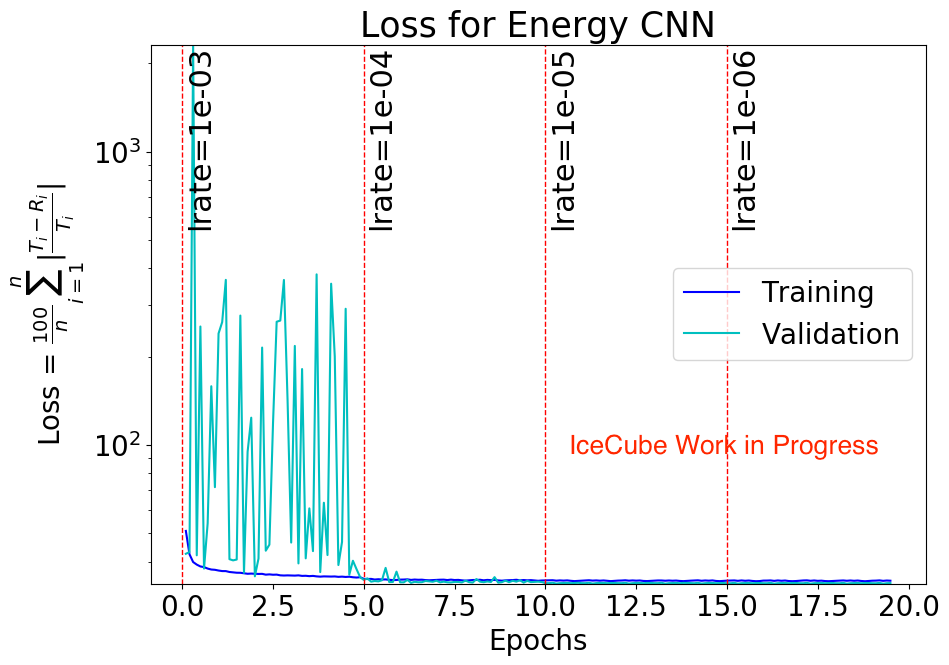}
\caption{Loss as a function of training progress. Each epoch is about 450,000 events and 10 epochs corresponds to one full pass through the data.}
\label{fig:Loss}}
\end{minipage}
\end{figure}

Figure \ref{fig:Loss} shows the progression of the training and validation loss as the network trains. Network training proceeds by minimizing the {\it loss}, defined here as the mean absolute percent error between the reconstructed and true energy of the event, chosen so that the CNN values accurately reconstructing low-energy as much as high-energy events. The loss function is iteratively minimized in discrete training ``epochs,'' or full passes through the training sample, until it reaches a plateau. Dropping the learning rate ({\it lrate}) in Figure \ref{fig:Loss}) helps with the instability of the validation sample while training. The performance of the CNN is evaluated after 11 epochs.

\section{Resolution and Timing Performance}

\begin{figure}[h]
\centering
\includegraphics[scale=0.3]{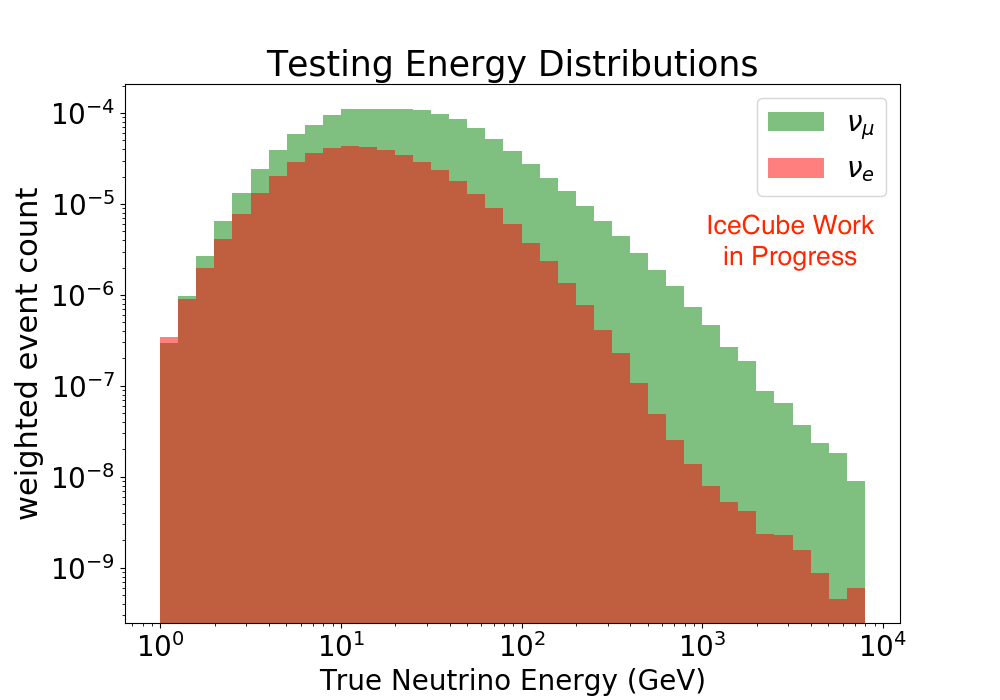}
\caption{The $\nu_\mu$ CC and $\nu_e$ CC testing samples: MC GENIE simulation with atmospheric flux weights applied to match data.}
\label{fig:TestingSamples}
\end{figure}

The CNN performance is evaluated on an independently generated GENIE neutrino sample with $\sim4$~million unweighted charge-current $\nu_\mu$ events and $\sim1.6$~million unweighted charge-current $\nu_e$ events. These distributions, weighted by the atmospheric flux, appear more like the expected energy distribution for data (Figures \ref{fig:TestingSamples}). This tests both the CNN's robustness to reconstruct a sample with a different distribution than its training sample and provides a more accurate estimation of the resolution for its potential to reconstruct low energy IceCube data. While the sample extends to 10 TeV, the performance evaluation will focus mostly on the 1-200 GeV region since the majority of the sample is expected at these energies and this is also the target region for an oscillation analysis.

\begin{figure}[h]
\center
\begin{minipage}{.4\textwidth}
\center{\includegraphics[scale=0.3]{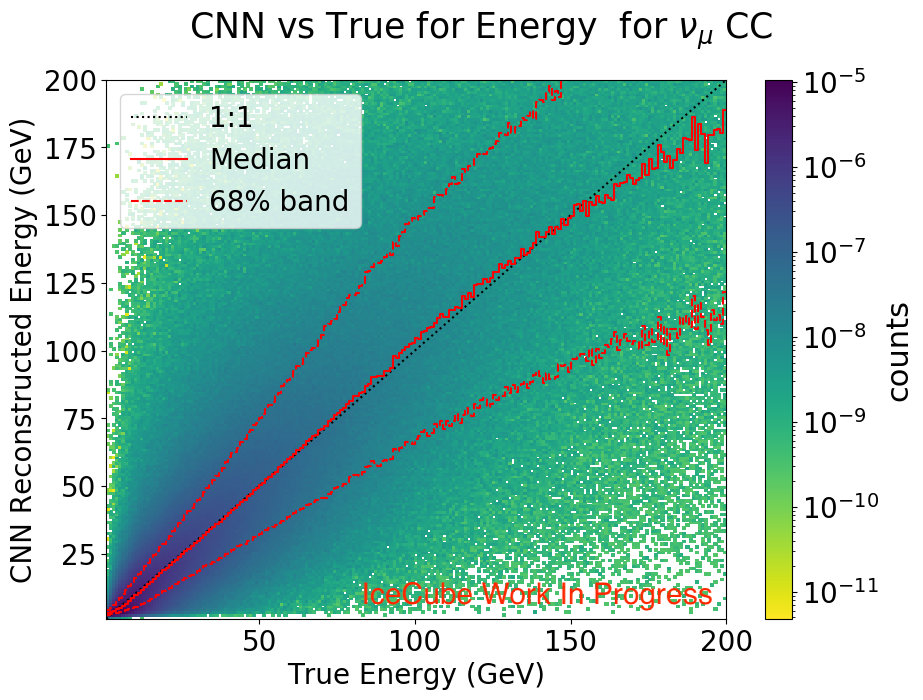}}
\end{minipage}\hspace{4em}%
\begin{minipage}{.4\textwidth}
 \center{\includegraphics[scale=0.3]{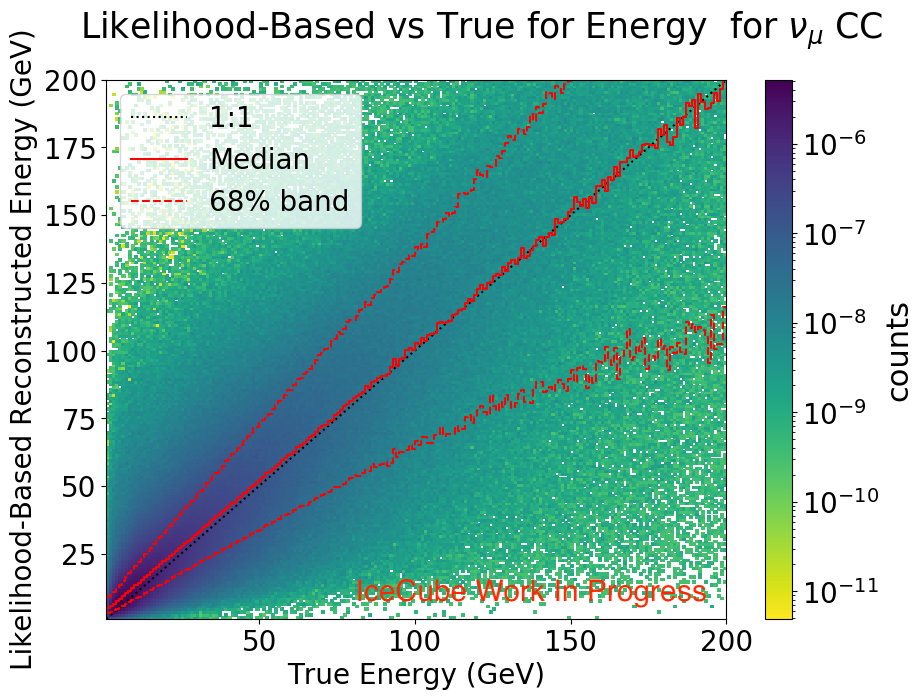}}
\end{minipage}
\caption{CNN (left) and likelihood-based (right) reconstructed vs true $\nu_\mu$ CC energy resolution. The median for each x-bin of true energy is shown in the solid red line with the 68\% band in the dotted red line.}
\label{fig:NuMu2D}
\end{figure}

Figure \ref{fig:NuMu2D} shows the CNN resolution compared to the true neutrino energy on the left, with the same comparison for IceCube's current likelihood-based reconstruction on the right. The CNN's median follows the ideal 1:1 line up to $\sim150$~GeV, then starts to slightly underestimate higher energy events. The likelihood-based reconstruction does not have the same underestimate, but the 68\% band is wider at higher energies. Improvements to the CNN are expected once more high energy training events are generated above the target region (200-500 GeV). Neural networks are known for having difficulty extrapolating data \cite{NN-extrapolate}; populating training data beyond the target region is one solution to improve resolution near the target boundary regions.

\begin{figure}[h]
\center
\begin{minipage}{.4\textwidth}
\center{\includegraphics[scale=0.28]{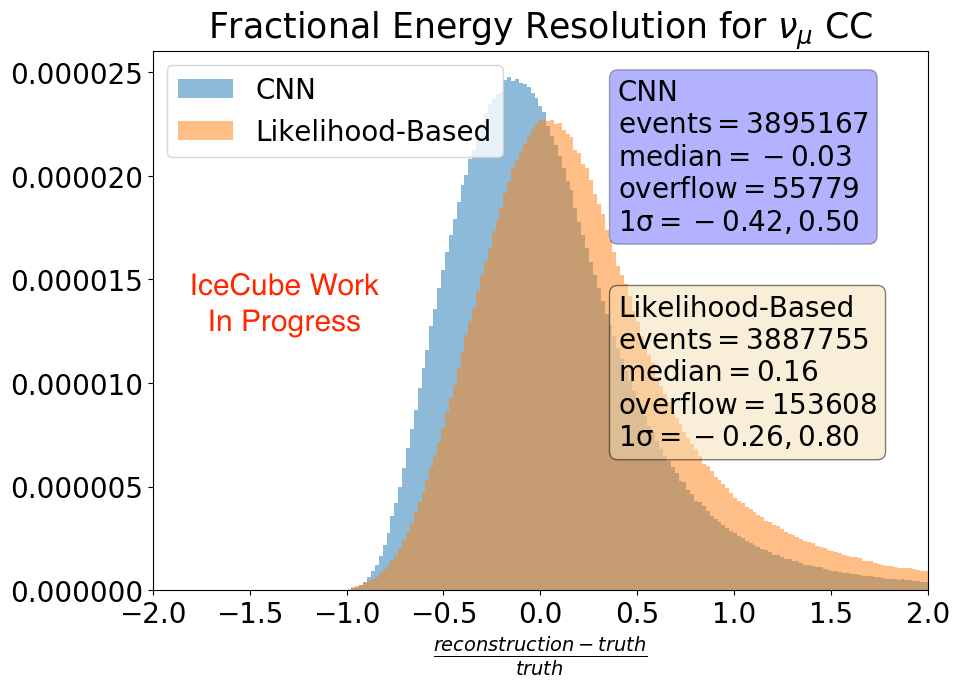}}
\caption{Fractional resolution of the reconstruction methods on the $\nu_\mu$ CC sample.}
\label{fig:NuMu_Resolution}\end{minipage}\hspace{4em}%
\begin{minipage}{.4\textwidth}
 \center{\includegraphics[scale=0.28]{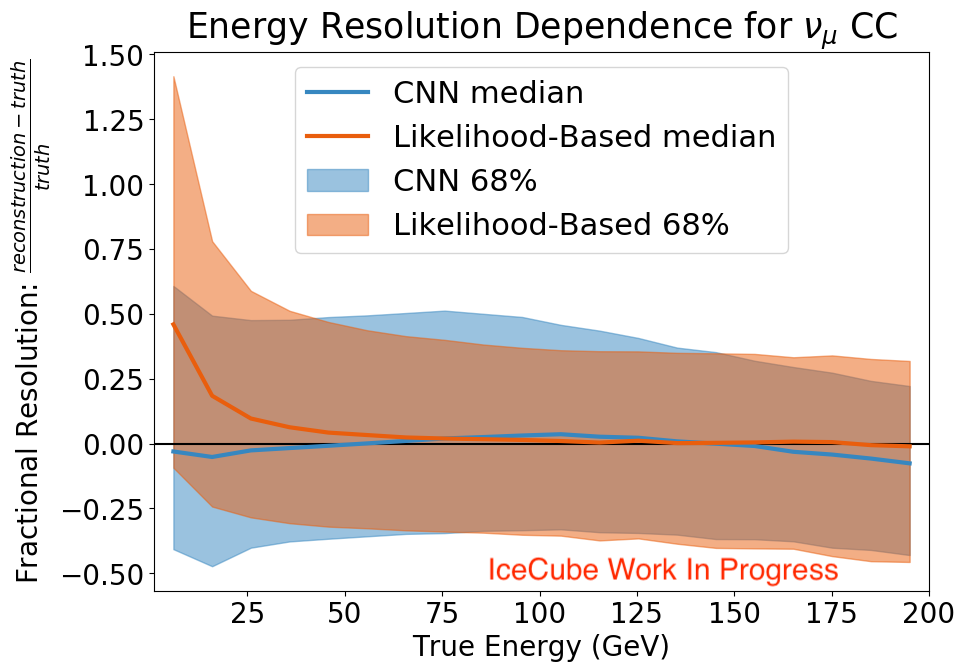}}
\caption{Fractional energy resolution as a function of the true energy on the $\nu_\mu$ CC sample.}
\label{fig:NuMu_Slices}
\end{minipage}
\end{figure}

Figures \ref{fig:NuMu_Resolution} and \ref{fig:NuMu_Slices} show a direct comparison between the CNN and likelihood-based energy reconstruction performance on the $\nu_\mu$ CC sample. Figure \ref{fig:NuMu_Resolution} shows the overall fractional resolution on the entire testing sample and Figure \ref{fig:NuMu_Slices} explores the fractional reconstruction as a function of the true energy. The CNN has an advantage at the lowest energies with its median bias offset near zero and 68\% band well resolved. This is particularly important since the expected energy resolution for the data will have the highest statistics in this low energy region. This explains the better overall 1$\sigma$ spread in Figure \ref{fig:NuMu_Resolution} for the CNN, since the $\nu_\mu$ CC performance sample is low-energy dominated.

\begin{figure}[h]
\center
\begin{minipage}{.4\textwidth}
\center{\includegraphics[scale=0.3]{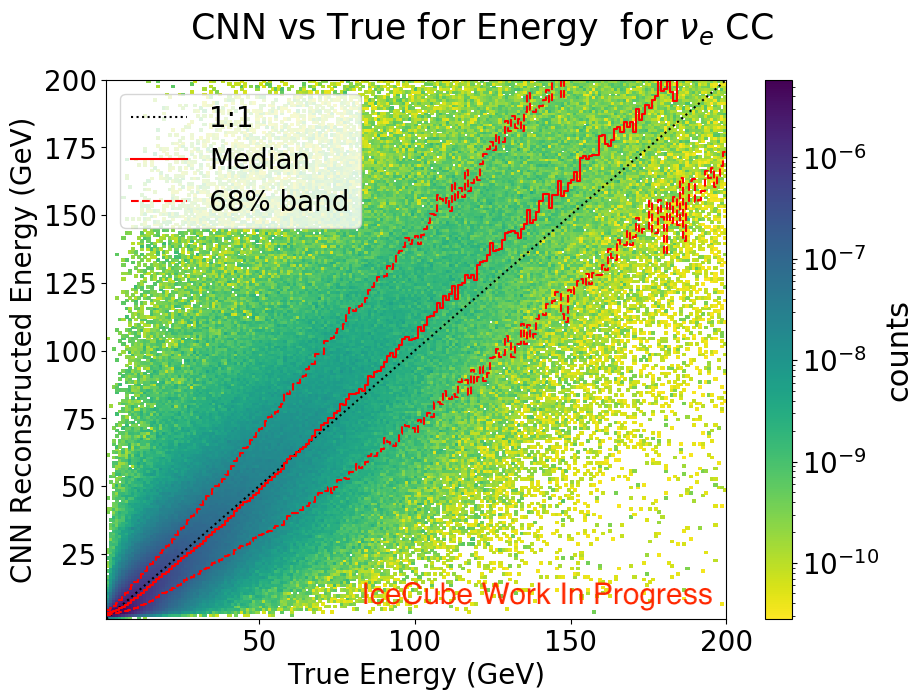}}
\end{minipage}\hspace{4em}%
\begin{minipage}{.4\textwidth}
 \center{\includegraphics[scale=0.3]{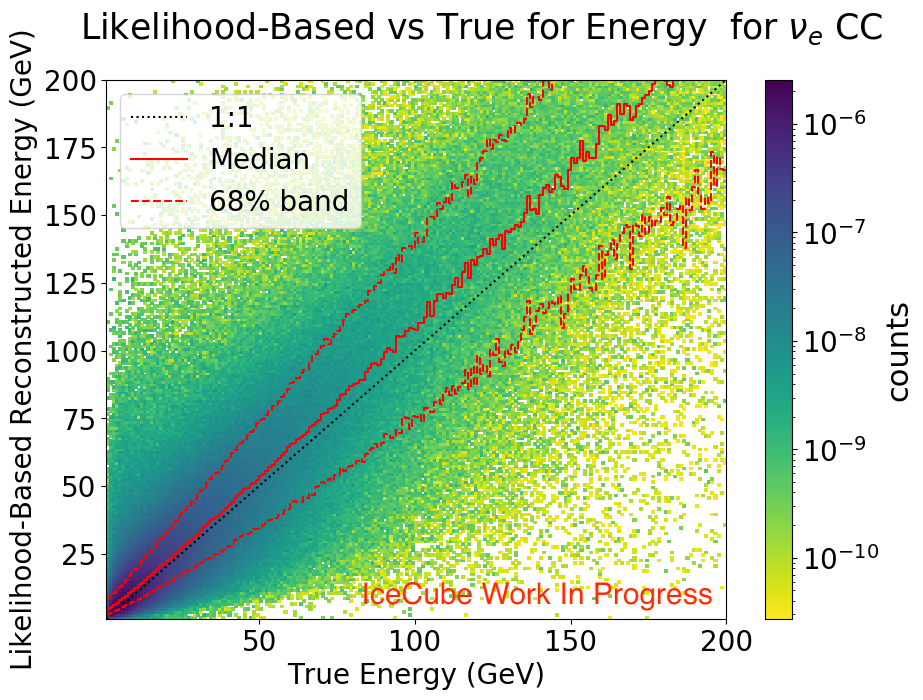}}
\end{minipage}
\caption{CNN (left) and likelihood-based (right) reconstructed vs true $\nu_e$ CC energy resolution. The median for each x-bin of true energy is shown in the solid red line with the 68\% band in the dotted red line.}
\label{fig:NuE2D}
\end{figure}

For the $\nu_e$ CC events, the CNN  slightly outperforms the likelihood-based reconstruction  at all energies (Figure \ref{fig:NuE2D}). Instead of the average underestimation of the $\nu_\mu$ CC energy in the 150-200 GeV region, the CNN has a slight overestimation in this region which matches the likelihood-based reconstruction's median trend. Since the network is trained on $\nu_\mu$ CC events only, it only sees the hadronic component of the cascade, which has less light per unit energy than in electromagnetic cascades. This could be the cause for the network to overestimate on average the $\nu_e$ CC events.

\begin{figure}[h]
\center
\begin{minipage}{.4\textwidth}
\center{\includegraphics[scale=0.28]{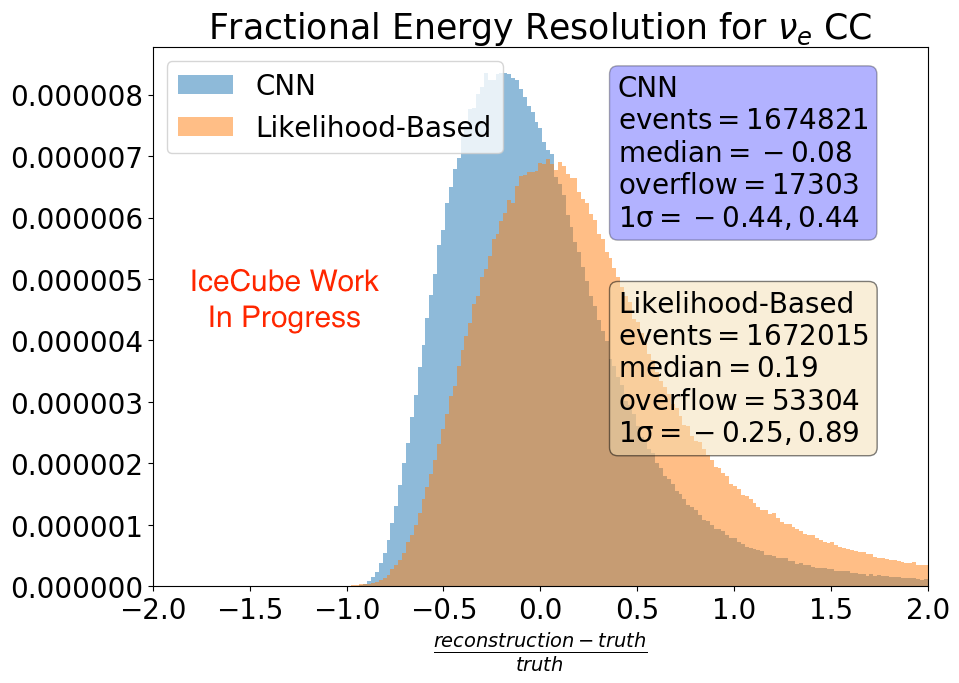}}
\caption{Fractional resolution of the reconstruction methods on the $\nu_e$ CC sample.}
\label{fig:NuE_Resolution}\end{minipage}\hspace{4em}%
\begin{minipage}{.4\textwidth}
 \center{\includegraphics[scale=0.28]{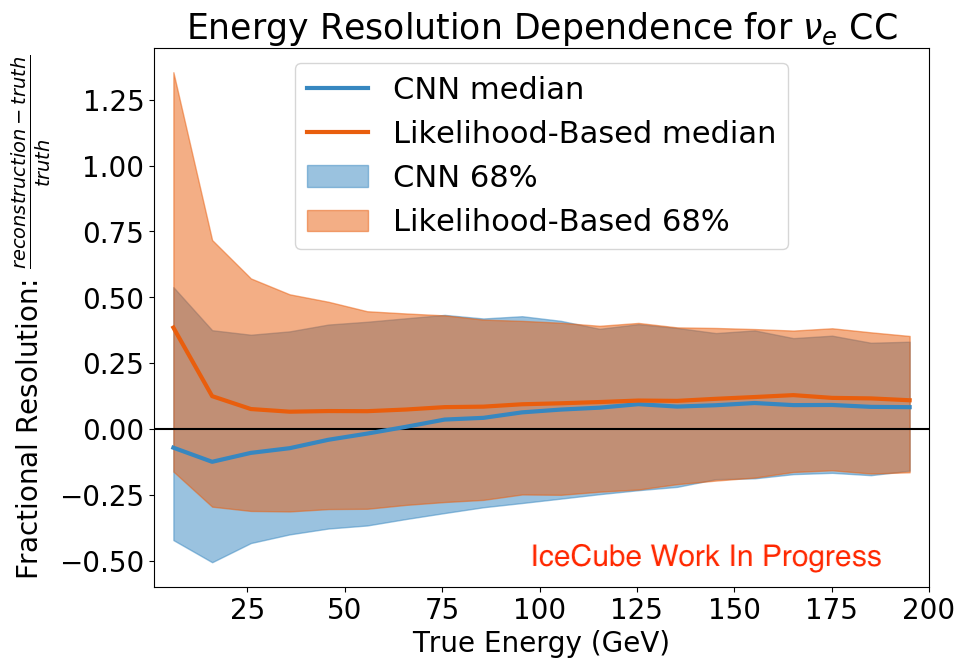}}
\caption{Fractional energy resolution as a function of the true energy on the $\nu_e$ CC sample.}
\label{fig:NuE_Slices}
\end{minipage}
\end{figure}

The direct comparisons for the CNN and likelihood-based energy reconstruction for the $\nu_e$ sample are shown in Figure \ref{fig:NuE_Resolution} and \ref{fig:NuE_Slices}. The 1$\sigma$ spread of the overall sample is better for the CNN (Figure \ref{fig:NuE_Resolution}). The CNN also has minimal offset bias for the fractional resolution at the lowest energies, improving on the likelihood-based estimation in this region (Figure \ref{fig:NuE_Slices}).

The reconstruction methods' runtimes are extremely important for the IceCube oscillations analysis, which expects a high-statistics sample ($\sim650$,000 neutrinos) collected over 10 years of livetime. Neural networks are known for their quick evaluation times, particularly when using GPUs. Table \ref{table:TimeTable} shows the timing for the CNN and the likelihood-based reconstructions, with the CNN evaluated both on a CPU and a GPU. Note that the likelihood-based method reconstructs a total of 8 variables simultaneously, while the CNN is currently only returning a single variable. Using the CNN to reconstruct more variables will only change the numbers quoted in Table \ref{table:TimeTable} by less than a factor of 10.

\begin{table}[htp]
\begin{minipage}{\textwidth}
\begin{center}
 \begin{tabular}{|c | c | c | c |} 
 \hline
 \textbf{Method} & \textbf{Average Time per Event (seconds)} & \textbf{Events per Day per Single Core} \\[0.5ex] 
 \hline
CNN on GPU & 0.0077 & 11,000,000 \\
\hline
CNN on CPU & 0.27 & 320,000 \\
\hline
Likelihood-Based & 40  & 2,100 \\
\hline
\end{tabular} \\
\caption{Comparing the time each method takes to reconstruct neutrino events.}
\label{table:TimeTable}
\end{center}
\end{minipage}
\end{table}

IceCube has access to computer clusters with both GPU and CPU nodes. Utilizing Michigan State University's High Performance Computing Cluster's GPUs yielded results for the performance set in about 15 minutes. The expected size of the data analysis set is estimated to be about 1/8 of the size of the performance MC set, so that reconstruction should be complete in less than 5 minutes. 

\section{Conclusion and Outlook}

Reconstructing the energy of 10-GeV scale IceCube neutrino events with a CNN provides comparable resolution to the current likelihood-based reconstruction but with much shorter runtimes. The low-energy CNN is still being optimized, in particular the network configuration and settings. Additional work is being conducted to explore the possibility of using a similar network structure to reconstruct the direction of the neutrino \cite{Shiqi-proceedings} and interaction vertex, along with classifications for particle identification and background rejection. Using the CNN for these purposes will provide an oscillation analysis with the necessary reconstructed variables and tools to apply cuts.

\section{Acknowledgements}

This material is based upon work supported by the National Science Foundation Graduate Research Fellowship Program under Grant No. DGE1848739 and NSF-1913607. Any opinions, findings, and conclusions or recommendations expressed in this material are those of the author(s) and do not necessarily reflect the views of the National Science Foundation. Additional acknowledgements included for the collaboration separately.

\bibliographystyle{ICRC}
\setlength{\bibsep}{3pt}
\bibliography{references}

\providecommand{\href}[2]{#2}\begingroup\raggedright\begin{thebibliography}{10}

\bibitem{SK-Osc}
{\bfseries Super-K} Collaboration, Y.~Fukuda {\em et~al.}
  \href{http://dx.doi.org/10.1103/PhysRevLett.81.1562}{{\em Phys. Rev. Lett.}
  {\bfseries 81} (Aug, 1998) 1562--1567}.

\bibitem{SNO-Osc}
{\bfseries SNO} Collaboration, Q.~R. Ahmad {\em et~al.}
  \href{http://dx.doi.org/10.1103/PhysRevLett.87.071301}{{\em Phys. Rev. Lett.}
  {\bfseries 87} (Jul, 2001) 071301}.

\bibitem{PhysRevLett.115.081102}
{\bfseries IceCube} Collaboration, M.~G. Aartsen {\em et~al.}
  \href{http://dx.doi.org/10.1103/PhysRevLett.115.081102}{{\em Phys. Rev.
  Lett.} {\bfseries 115} (Aug, 2015) 081102}.

\bibitem{Science-TXS}
M.~G. Aartsen {\em et~al.}
  \href{http://dx.doi.org/10.1126/science.aat2890}{{\em Science} {\bfseries
  361} no.~6398, (2018) 147--151}.

\bibitem{Aartsen:2016nxy}
{\bfseries IceCube} Collaboration, M.~G. Aartsen {\em et~al.}
  \href{http://dx.doi.org/10.1088/1748-0221/12/03/P03012}{{\em JINST}
  {\bfseries 12} no.~03, (2017) P03012}.

\bibitem{DeepCore}
R.~Abbasi {\em et~al.}
  \href{http://dx.doi.org/https://doi.org/10.1016/j.astropartphys.2012.01.004}{{\em
  Astroparticle Physics} {\bfseries 35} no.~10, (2012) 615--624}.

\bibitem{IceCube-Osc}
{\bfseries IceCube} Collaboration, M.~G. Aartsen {\em et~al.}
  \href{http://dx.doi.org/10.1103/PhysRevLett.120.071801}{{\em Phys. Rev.
  Lett.} {\bfseries 120} (Feb, 2018) 071801}.

\bibitem{Krizhevsky:cnn}
A.~Krizhevsky, I.~Sutskever, and G.~Hinton
  \href{http://dx.doi.org/10.1145/3065386}{{\em NeurIPS} {\bfseries 25} (01,
  2012) }.

\bibitem{kaudererabrams2017quantifying}
E.~Kauderer-Abrams, ``Quantifying translation-invariance in convolutional
  neural networks,'' 2017.

\bibitem{Huennefeld:2017tT}
{\bfseries IceCube} Collaboration
  \href{http://dx.doi.org/10.22323/1.301.1057}{{\em PoS} {\bfseries ICRC2017}
  (2017) 1057}.

\bibitem{GENIE}
C.~Andreopoulos {\em et~al.}
  \href{http://dx.doi.org/https://doi.org/10.1016/j.nima.2009.12.009}{{\em
  Nucl. Instrum. Methods Phys. Res. A} {\bfseries 614} no.~1, (2010) 87--104}.

\bibitem{NN-extrapolate}
D.~C. Psichogios and L.~H. Ungar
  \href{http://dx.doi.org/https://doi.org/10.1002/aic.690381003}{{\em AIChE
  Journal} {\bfseries 38} no.~10, (1992) 1499--1511}.

\bibitem{Shiqi-proceedings}
{\bfseries IceCube} Collaboration {\em PoS} {\bfseries ICRC2021} (these
  proceedings) 1054.

\end{thebibliography}\endgroup


\clearpage
\section*{Full Author List: IceCube Collaboration}




\scriptsize
\noindent
R. Abbasi$^{17}$,
M. Ackermann$^{59}$,
J. Adams$^{18}$,
J. A. Aguilar$^{12}$,
M. Ahlers$^{22}$,
M. Ahrens$^{50}$,
C. Alispach$^{28}$,
A. A. Alves Jr.$^{31}$,
N. M. Amin$^{42}$,
R. An$^{14}$
% ...
\\

\noindent
% ...
$^{12}$ Universit{\'e} Libre de Bruxelles, Science Faculty CP230, B-1050 Brussels, Belgium \\
% ...
$^{14}$ Department of Physics and Laboratory for Particle Physics and Cosmology, Harvard University, Cambridge, MA 02138, USA \\
% ...
$^{17}$ Department of Physics, Loyola University Chicago, Chicago, IL 60660, USA \\
$^{18}$ Dept. of Physics and Astronomy, University of Canterbury, Private Bag 4800, Christchurch, New Zealand \\
% ...
$^{22}$ Niels Bohr Institute, University of Copenhagen, DK-2100 Copenhagen, Denmark \\
% ...
$^{28}$ D{\'e}partement de physique nucl{\'e}aire et corpusculaire, Universit{\'e} de Gen{\`e}ve, CH-1211 Gen{\`e}ve, Switzerland \\
% ...
$^{31}$ Karlsruhe Institute of Technology, Institute for Astroparticle Physics, D-76021 Karlsruhe, Germany  \\
% ...
$^{42}$ Bartol Research Institute and Dept. of Physics and Astronomy, University of Delaware, Newark, DE 19716, USA \\
% ...
$^{50}$ Oskar Klein Centre and Dept. of Physics, Stockholm University, SE-10691 Stockholm, Sweden \\
% ...
$^{59}$ DESY, D-15738 Zeuthen, Germany \\
% ...

\subsection*{Acknowledgements}

\noindent
USA {\textendash} U.S. National Science Foundation-Office of Polar Programs,
U.S. National Science Foundation-Physics Division,
U.S. National Science Foundation-EPSCoR,
Wisconsin Alumni Research Foundation,
Center for High Throughput Computing (CHTC) at the University of Wisconsin{\textendash}Madison,
Open Science Grid (OSG),
Extreme Science and Engineering Discovery Environment (XSEDE),
Frontera computing project at the Texas Advanced Computing Center,
U.S. Department of Energy-National Energy Research Scientific Computing Center,
Particle astrophysics research computing center at the University of Maryland,
Institute for Cyber-Enabled Research at Michigan State University,
and Astroparticle physics computational facility at Marquette University;
Belgium {\textendash} Funds for Scientific Research (FRS-FNRS and FWO),
FWO Odysseus and Big Science programmes,
and Belgian Federal Science Policy Office (Belspo);
Germany {\textendash} Bundesministerium f{\"u}r Bildung und Forschung (BMBF),
Deutsche Forschungsgemeinschaft (DFG),
Helmholtz Alliance for Astroparticle Physics (HAP),
Initiative and Networking Fund of the Helmholtz Association,
Deutsches Elektronen Synchrotron (DESY),
and High Performance Computing cluster of the RWTH Aachen;
Sweden {\textendash} Swedish Research Council,
Swedish Polar Research Secretariat,
Swedish National Infrastructure for Computing (SNIC),
and Knut and Alice Wallenberg Foundation;
Australia {\textendash} Australian Research Council;
Canada {\textendash} Natural Sciences and Engineering Research Council of Canada,
Calcul Qu{\'e}bec, Compute Ontario, Canada Foundation for Innovation, WestGrid, and Compute Canada;
Denmark {\textendash} Villum Fonden and Carlsberg Foundation;
New Zealand {\textendash} Marsden Fund;
Japan {\textendash} Japan Society for Promotion of Science (JSPS)
and Institute for Global Prominent Research (IGPR) of Chiba University;
Korea {\textendash} National Research Foundation of Korea (NRF);
Switzerland {\textendash} Swiss National Science Foundation (SNSF);
United Kingdom {\textendash} Department of Physics, University of Oxford.

\end{document}